\documentclass[traditabstract]{aa} 
\usepackage{graphicx}
\usepackage{natbib}
\usepackage{mathabx}
\usepackage{deluxetable}

\usepackage[pdftex,                
bookmarks=true,                    
bookmarksnumbered=true,            
colorlinks=true,            
citecolor=blue,         
linkcolor=blue,     
menucolor=blue,         
urlcolor=cyan,       
linkbordercolor={0 0 1},           
pdfborder= {0 0 1},
frenchlinks=True]{hyperref}
\usepackage{txfonts}
\providecommand{\eprint}[1]{\href{http://arxiv.org/abs/#1}{#1}}
\providecommand{\adsurl}[1]{\href{#1}{ADS}}
\providecommand{\figwidth}{0.5\textwidth}

\def\micron{\ensuremath{\mu m}}

\newcommand{\rearth}{\ensuremath{R_\Earth}}
\newcommand{\mearth}{\ensuremath{M_\Earth}}
\newcommand{\verysmallp}{\ensuremath{1-2\,\rearth}}
\newcommand{\smallp}{\ensuremath{2-4\,\rearth}}
\newcommand{\mediump}{\ensuremath{4-8\,\rearth}}
\newcommand{\largep}{\ensuremath{8-32\,\rearth}}

\begin{document}

\title{On High-Contrast Characterization of Nearby, Short-Period
  Exoplanets with Giant Segmented-Mirror Telescopes}

\authorrunning{Crossfield}
\titlerunning{GSMT High-Contrast Characterization of Short-Period Exoplanets}

   \author{Ian J. M. Crossfield
             }

          \institute{
             Max-Planck Instit\"ut f\"ur Astronomie, K\"onigstuhl 17, 69117, Heidelberg, Germany \\
            \email{\href{mailto:ianc@mpia.de}{ianc@mpia.de}}
            }

   \date{A\&A Accepted, 24.\ Jan 2013.}

 
  \abstract
  {Measurements of the frequency with which short-period planets occur
    around main sequence stars allows a direct prediction of the
    number and types of such planets that will be amenable to
    characterization by high-contrast instruments on future giant
    segmented-mirror telescopes (GSMTs).  Adopting conservative
    assumptions, I predict of order 10 planets with radii
    $R_P=1-8\,\rearth$ and equilibrium temperatures $\lesssim400$~K
    should be accessible around stars within 8~pc of the Sun. These
    numbers are roughly the same for both near-infrared observations
    of scattered starlight and mid-infrared observations of planetary
    thermal emission, with the latter observations demonstrating
    greater relative sensitivity to smaller and cooler planets.
    Adopting the conservative assumption that planets with
    $R_P=\verysmallp$ and \smallp\ occur with equal frequency, I
    predict a 40\% chance that a planet with $R_P=\verysmallp$ and
    equilibrium temperature 200--250~K will accessible to
    high-contrast thermal infrared characterization; this would be a
    compelling object for further study. To validate these
    predictions, more detailed analyses are needed of the occurrence
    frequencies of low-mass planets around M dwarfs, both in the {\em
      Kepler} field and in the solar neighborhood.  Several planets
    already discovered by radial velocity surveys will be accessible
    to near-infrared high-contrast GSMT observations, including the
    low-mass planets $\alpha$~Cen~Bb and (depending on their albedos)
    GJ~139c and~d, GJ~876b and~c, {and $\tau$~Cet~b, c, and~d;
      $\tau$~Cet~f} would be amenable to thermal infrared
    characterization.  Further efforts to model the near-infrared
    reflectance and mid-infrared emission of these and other
    short-period planets are clearly warranted, and will pave the way
    for the interpretation of future high-contrast characterization of
    a variety of planets around the nearest stars.

}
   { }
   { }
   { }
   {}

   \keywords{Planets and satellites: general --- Planets and
     satellites: detection --- Instrumentation: adaptive optics ---
      Techniques: high angular resolution --- Methods: numerical}

   \maketitle
%

 \section{Introduction}
A 1.1~Earth-mass ($\mearth$) planet {candidate} has been announced
around $\alpha$~Centauri~B \citep{dumusque:2012}, {and a family of five
candidates with minimum masses 2--10\,\mearth\ has been announced
around $\tau$~Ceti \citep{tuomi:2013}. If confirmed, the discovery of
so many low-mass planets around stars very near the Sun} would be a
strong sign that such planets are common. Such a conclusion would be
consistent with early results from the Kepler mission, which indicate
that the frequency of planet candidates increases sharply toward
smaller planetary radii ($R_P \lesssim 4\,\rearth$) for orbital
periods $P<250$\,d \citep{howard:2012,dong:2013,fressin:2013}.  These
Kepler results also suggest a dramatic increase in the frequency of
small planets (radii $R_P=\smallp$) orbiting cooler stars
\citep{howard:2012}. Because such stars are more numerous than
sun-like stars, there should be many such planets in the Solar
neighborhood. Because most of these planets orbit M dwarfs, their
equilibrium temperatures $T_{\rm eq}$ are quite cool: typically
200--400\,K.

The prospect of numerous temperate, small, low-mass planets orbiting
nearby stars is an exciting one. Such planets may lie in their host
stars' ``habitable zones'' and be capable of supporting liquid water
on any solid surface
\citep{huang:1959,kasting:1993,pierrehumbert:2011}.  These planets
around nearby stars are also the only ones for which {\em in situ}
exploration could be even remotely feasible in the foreseeable future
\citep[e.g.,][and references therein]{crawford:2011}. For now, the
proximity of nearby systems and their attendant high photon fluxes
make these planets the most amenable to remote characterization of
their atmospheric and/or surface compositions.

In recent years much effort has been devoted to preparing for
observations of potential Earth analogues transiting cool M dwarf
using, e.g., JWST \citep[][]{deming:2009,kaltenegger:2009}. However,
because even short period planets are unlikely to transit, the nearest
accessible transiting planets will be around rather fainter, more
distant stars. The durations of these transits will be short and their
signal to noise low, so tens or hundreds of observations will be
necessary for significant atmospheric measurements. Phase curve
observations provide higher duty cycles and may be able to
characterize small, non-transiting planets around stars closer to the
Sun \citep{selsis:2011,maurin:2012}. Spacecraft capable of making such
measurements have been proposed
\citep[e.g.,][]{vasisht:2008,swain:2010finesse,tinetti:2012}, but
these observations will require extremely stringent spectrophotometric
stability (better than $10^{-4}$ over periods of days). Considering
these various limitations, it is reasonable to explore alternative
modes of exoplanetary characterization.

In this paper I present the first quantitative estimates of the local
exoplanet population accessible to ground-based high-contrast
observations, and a discussion of how such observations can directly
constrain the demographics of the local planet population.  Though it
has been appreciated for at least a decade that the next generation of
ground-based giant segmented-mirror telescopes (GSMTs) can
characterize small, cool planets via such observations
\citep{angel:2003,hawarden:2003}, Kepler's ability to measure planet
frequency as a function of $P$, $R_P$, and stellar effective
temperature $T_{\rm eff}$ with unprecedented precision permits the
first informed estimate of the number of short-period planets that
will be accessible to future large-aperture facilities.

Sec.~\ref{sec:methods} discusses simulations of statistically
representative planetary systems around nearby stars and an assessment
of high-contrast GSMT instruments' ability to characterize these
planets.  Sec.~\ref{sec:results} then presents the number of planets
accessible to such characterization under various assumptions of
planetary occurrence frequency and instrumental capabilities, and
discusses the consequences of these assumptions. The main results are
summarized in Table~\ref{tab:results}. In brief, of order 10 planets
are predicted to be amenable to such characterization at either
near-infrared (NIR, $<2.5\,\micron$) or mid-infrared (MIR,
$\sim10\,\micron$) wavelengths, and this number varies by no more than
a factor of 2 under rather more optimistic or pessimistic
assumptions. In addition, there is significant likelihood (at least
40\%, and probably $\sim$4$\times$ higher) that such observations will
be able to characterize an Earth or Venus analogue ($R_P=\verysmallp$,
$T_{\rm eq}=200-250$~K). Finally, I summarize the results of this work
and suggest useful future work in Sec.~\ref{sec:conclusion}.

\section{Methods}
\label{sec:methods}
\subsection{Defining the Local, Observable Stellar Population}
To construct an observable sample of nearby stars, I take the 8\,pc
sample of \cite{kirkpatrick:2012} and remove all white dwarfs and
spectral types later than M7 (to avoid brown dwarfs). White and brown
dwarfs may have low mass planets, and these planets could even be
temperate \citep{andreeschev:2004,monteiro:2010}, but at present the
frequency of planets in such systems remains very poorly
constrained. Because high-contrast observations have difficulty
attaining the contrast necessary for planet characterization in the
presence of nearby bright point sources \citep[mitigation strategies
exist but require customized instruments; see e.g.,
][]{crepp:2010,cady:2011}, I remove all known binaries with apparent
separations $\lesssim5$''. Stars with fainter companions (e.g.,
Procyon~A and Sirius~A) are retained. For objects listed by
\cite{kirkpatrick:2012} with a range of spectral types, I take the
mean of the specified range. Some M dwarfs are listed without any
spectral type, and for these I take the spectral types specified by
\cite{rojas-ayala:2012}.  The final list contains 92, 20, 6, 2, and 4
stars of types M, K, G, F, and A, respectively.

Accurate stellar parameters are not known for all these objects, so I
adopt a homogeneous approach and assign $T_{\rm eff}$ and stellar
radii using empirical, interferometrically-determined values
\citep{boyajian:2012a,boyajian:2012b} for each spectral type. Apparent
V-band magnitudes are taken from the UCAC4 Catalogue
\citep{zacharias:2012}, converted to absolute V magnitudes using the
parallax values in \cite{kirkpatrick:2012}, and converted to stellar
masses using the mass-luminosity relationship of \cite{henry:1993};
{using the K band relationship of \cite{delfosse:2000} affects the
  masses thus derived by roughly 15\%, but this does not significantly
  affect my results.} Finally, a BT-Settl stellar model
\citep{allard:2011} is generated by interpolating in $\log g$ and
$T_{\rm eff}$ to the stellar parameters, and accounting for the star's
size and distance from the Sun.

Note that this 8~pc sample is sufficiently small that high-quality
spectra could be obtained for all these targets with relative ease.
Spectra at low resolution would provide a useful absolute flux
calibration, and spectra at high resolution would permit stellar
parameters and abundances to be derived at high precision. Such a
homogeneous stellar characterization effort would be highly desirable,
and would reduce the need for the assumptions about stellar parameters
described above.

\subsection{Modeling the Local Exoplanet Population}
\label{sec:pop}
I estimate the distribution of the local exoplanet population using an
analysis of planet frequency as a function of $T_{\rm eff}$, $P$, and
$R_P$ from the Kepler planet candidate sample derived from {the first
three quarters of Kepler observations
\citep{borucki:2011,howard:2012}. Although these frequencies are
computed for planet candidates and not fully validated planets, the
occurrence of false positives in the sample is below 15\%
\citep{morton:2011,fressin:2013}.  Furthermore, planet candidates with
radii $R_P=2-4\,\rearth$ (the targets most accessible to high-contrast
characterization) have a contamination fraction of $<7\%$
\citep{fressin:2013}. Thus the Kepler false positive rate has at most
a minor effect on the results presented here.}

Hot Jupiters are only half as frequent in the Kepler sample as
observed by radial velocity surveys \citep{howard:2012}. The authors
attribute this discrepancy to the lower metallicity of the Kepler
sample, under which conditions fewer massive planets should be
expected \citep{santos:2004,fischer:2005}. If this discrepancy is
confirmed, then the simulations presented here underestimate the
frequency of large planets by as much as a factor of 2; these planets
are intrinsically rare, so this is a small effect.  The frequency of
Neptune-mass and smaller planets depends less strongly on stellar
metallicity than does the frequency of Jupiter-mass objects
\citep{sousa:2008,buchhave:2012}; since I show below that smaller
planets comprise the bulk of the planet population accessible to
high-contrast characterization, differences between the Kepler and
local stellar populations should not strongly affect these results
(this assumption will be tested in the future by high-precision radial
velocity surveys).

I begin by determining the frequency of planets of given $R_P$ for
each star, based on its $T_{\rm eff}$.  I assume that binary stars
have planet frequencies consistent with those of single stars, and
that \mediump\ and \largep\ planets occur with frequencies of $1.7\%$
and $0.79\%$, respectively \citep[see Figure~8 of][]{howard:2012}.
For \smallp\ planets, I compute a frequency using Eq.~9 of
\cite{howard:2012}, which was defined using $T_{\rm eff}$ in the range
3600--7100~K. I extend this relation to lower and higher $T_{\rm eff}$
by assuming that beyond the specified range the frequency distribution
of planets is uniform and everywhere continuous.  In
Sec.~\ref{sec:astroparam} I quantify the effect of this
assumption. \cite{howard:2012}'s planet statistics are incomplete for
$R_P < 2 \rearth$, but their results suggest \verysmallp\ planets are
at least as common as \smallp\ planets.  I conservatively assume that
planets in these two size ranges occur with equal frequency,
{consistent with more recent analyses \citep{dong:2013,fressin:2013}.}
Throughout this work, I assume a uniform distribution of planet sizes
within each of the $R_P$ ranges noted above.

The preceding steps determine the number of planets of varying $R_P$
expected around a given star for $P<50$~d.  In each of $10^4$ trials I
assign to each star either no planet or a planet in one of the four
$R_P$ bins listed above, with the appropriate frequencies. I then
assume that $P$ in each radius range follows that given by Eq.~8 and
Table~5 of \cite{howard:2012}. Though their planet population
statistics are limited to $P<50$~d, even over this small range the
frequency of planets increases in $\ln P$ space for $P>10$~d. I
conservatively assume that planet frequency is flat for $P>50$\,d
\citep[see also][]{dong:2013}. Letting planet frequency increase
toward longer periods only slightly increases the number of accessible
planets, because the planets most accessible to high-contrast
characterization lie on short-period orbits (see
Sec.~\ref{sec:astroparam}).  Future studies should characterize planet
frequency at longer orbital periods (this is a primary goal of the
Kepler mission) and should explore the dependence of planet frequency
on $T_{\rm eff}$ in greater detail.

The result of this process is a planet population that provides a
direct estimate of the frequency of planets with $R_P$ and $P$ around
each star in the 8~pc sample. Planetary systems may be more diverse
than predicted by this simple model: for example, the M dwarf GJ~876
has multiple massive planets \citep[e.g.,][]{baluev:2011} while
short-period planets of even very low mass have been excluded around
the low-mass Barnard's Star and Proxima Centauri
\citep{endl:2008,zechmeister:2009,choi:2012}. Nonetheless the modeling
approach described above has the advantages of being relatively simple
and of providing estimates across a broad range of planetary and
stellar parameter space that decently approximate the planet
demographics presented by \cite{howard:2012}.  Future efforts should
strive to quantify planet frequency over as broad a range as possible
of $T_{\rm eff}$, $P$, and $R_P$ to obviate the need for some of the
assumptions made above.

\subsection{Estimating High-Contrast Instrument Performance}
\label{sec:ao}
Adaptive optics (AO) systems designed to provide very high image
quality are often termed ``high contrast'' systems, a name which
emphasizes their sensitivity to faint objects located near much
brighter ones; see \cite{oppenheimer:2009} and \cite{mawet:2012} for
recent reviews on this topic.  At least one high-contrast system is
already operating on-sky at the 5~m Hale telescope and reaching
contrast levels of $\sim 10^{-7}$ \citep{oppenheimer:2012}, and even
more capable instruments are under construction for existing 8\,m telescopes
\citep{macintosh:2006gpi, beuzit:2006}.  End-to-end simulations and
testing of these latter instruments suggest they may reach final
contrasts of roughly $10^{-8}$
\citep{mesa:2011,thomas:2011,wildi:2011}. At this level of
performance, short-period exoplanets will likely remain beyond the
reach of such instruments (see Sec.~\ref{sec:angular}).

The next generation of GSMTs will have diameters ranging from 25\,m
\citep[GMT;][]{johns:2008} to 30\,m \citep[TMT;][]{nelson:2008} to
39\,m \citep[E-ELT;][]{gilmozzi:2007}, and all are projected to rely
heavily on AO-assisted, diffraction-limited observations to maximize
the science gains of these new facilities.  High-contrast NIR
instruments have been proposed and studied for all these telescopes
\citep{macintosh:2006pfi,matsuo:2010,brandl:2010,kasper:2010}. These
initial studies suggest that with moderate and reasonable
technological advances, such instruments will be sensitive to objects
at separations and contrast levels unprecedented for optical/infrared
telescopes.

High contrast instruments generally require highly specialized designs
to achieve their goals, and their performance varies across different
targets, telescopes, and wavelengths. I estimate the performance of a
generic GSMT high-contrast instrument using the analytic relations of
\cite{guyon:2005} and adopting the parameters in Table~4 of that work
unless noted otherwise (variations of instrument parameters are
discussed in Sec.~\ref{sec:obsparam})\footnote{Note that Eqs. 16 and
  17 of \cite{guyon:2005} contain an error: the coefficient 0.484
  should be 0.0484 (O.~Guyon, private communication, Jan~2013).}.  The
instrument is assumed to correct both phase and amplitude errors,
sensed via a Mach-Zehnder pupil plane wavefront sensor. The wavefront
sensing and science wavelengths are assumed to be identical to
minimize chromatic contrast errors \citep{guyon:2005}.  {The relations
  used compute raw contrast and not the effective contrast limits for
  detection;} I therefore assume that observational and
post-processing techniques can suppress the instantaneous PSF contrast
by a factor of 30 at all angular separations; current techniques
achieve suppression factors of 20 to several hundred at separations as
small as 4.5$\lambda/D$ \citep{marois:2008,crepp:2011,vogt:2011}, so
the value assumed here may be somewhat conservative. Finally,
following \cite{lunine:2008} I impose an absolute contrast floor of
$5\times10^{-9}$ in all cases.  The result is a simulation of
high-contrast performance with the phenomenologically correct
dependencies on wavelength, telescope size, stellar flux, and other
parameters\footnote{A Python software module used to calculate these
  quantities is available from the author's website, located (as of
  publication) at \url{http://www.mpia.de/homes/ianc/} }.

In these performance estimates I also crudely account for the effect
of finite stellar sizes, which are significant for several of the
nearest and largest stars.  For example, both $\alpha$~Cen~A and
Sirius~A have angular diameters $\theta=8$~mas, comparable to the
diffraction-limited angular resolution ($\lambda/D$) of a 30\,m
telescope observing at 1.2\,\micron. Resolved sources cause the
effective throughput of high-contrast observations to decrease
markedly at the smallest separations \citep{guyon:2006}, though the
attainable contrast is not adversely affected for proper choice of
coronagraph or other starlight suppression system
\citep{crepp:2009}. I approximate the deleterious effects of finite
stellar size by imposing a minimum inner working angle (IWA) of $(2.2
+ 0.4 \log_{10} [\theta (\lambda/D)^{-1}] ) \lambda/D$, a relation
which approximates the 50\% throughput limit of an ideal coronagraph
operating at $10^{-10}$ contrast \citep{guyon:2006}. This minimum
separation increases for larger $\theta$, more extreme contrast
levels, and less optimal coronagraph designs.  In this work finite
stellar size only affects a few star systems (those with the largest
$\theta$), and only in the $2\lambda/D$ IWA case described in
Sec.~\ref{sec:angular}. Future high-contrast observations probing near
or within $1\lambda/D$ \citep[][]{serabyn:2010} should consider these
effects in greater detail.

\subsection{Estimating the Observability of Short-Period Planets}
\label{sec:accessibility}
I assess the accessibility of each simulated planet to high-contrast
observations by considering both scattered starlight and thermal
emission from the planet. I assume that all planets will be observed
at quadrature, since radial velocity measurements will allow optimal
phasing of most interesting targets. I further assume that all planets
are on circular orbits and exhibit Lambertian scattering profiles, and
that the observatory can observe all stars in the target sample
(under currently published plans GSMTs will be located in both
hemispheres).

I assign a Bond and wavelength-independent geometric albedo ($A_B$ and
$A_g$, respectively) to each planet by drawing a value from a uniform
random distribution spanning 0.0--0.4, a range which is a roughly
representative for solar system planets \citep[see also][]{hu:2012}.
Although $A_B$ and $A_g$ describe different physical quantities
\citep[$A_g$ determines the strength of scattering at a given
wavelength, while $A_B$ determines a planet's equilibrium temperature
$T_{\rm eq}$, and thus its thermal blackbody emission; see
e.g.,][]{seager:2010}, for simplicity I use the same value for each
planet's $A_B$ and $A_g$.  This equality holds to within a factor of
two for planets in the Solar System. At NIR wavelengths most
observable planets in the simulations below are only accessible in
reflected light, so higher or lower albedos will respectively increase
or decrease the size of the accessible planet population; the effect
is reversed for observations at thermal infrared wavelengths. To each
planet I also assign a heat redistribution factor \citep[$f$; ][]{seager:2010} drawn from a uniform random distribution
spanning the range (0.25, 0.66), where the extrema of this range
respectively indicate zero and full redistribution of the incident
stellar flux. The redistribution factor does not significantly affect
the detectability of planets in reflected light, but would affect the
thermal infrared observations described in Sec.~\ref{sec:metis}.

Using $A_B$, $f$, and the physical system parameters I then compute
$T_{\rm eq}$ for each planet, and I assume that the planet radiates as
a blackbody with this temperature.  Finally, the planet to star flux
contrast in a given bandpass is the ratio of the sum of the planet's
reflected and thermal photon flux to the photon flux from the host
star. The planet is assumed to be observable if its contrast lies
above the sensitivity limits computed in Sec.~\ref{sec:ao}.

\section{Simulation Results}
\label{sec:results}
In this section I discuss the local population of planets accessible
to characterization by high-contrast observations under a number of
different assumptions about planet population parameters and
instrumental configurations.  I begin by presenting a set of baseline
astrophysical and observational parameters in
Sec.~\ref{sec:baseline}. Sec.~\ref{sec:astroparam} discusses the
effects of varying the parameters of the local planetary population,
and Sec.~\ref{sec:obsparam} discusses the impact of various telescope
and instrument design choices. For all cases discussed, the number of
planets of various $R_P$ and $T_{\rm eq}$ accessible to high-contrast
characterization is listed in Table~\ref{tab:results}.

\subsection{Baseline Scenario}
\label{sec:baseline}
The baseline observational parameters are a 30\,m telescope capable of
observing down to an IWA of $3\lambda/D$: this instrument is analogous
to TMT's PFI or SEIT instruments
\citep{macintosh:2006pfi,matsuo:2010}, but with better wavefront
sensor performance. This baseline case provides an angular resolution
midway between the 25\,m GMT and 39\,m E-ELT.  I adopt a baseline
observing wavelength of 1.2\,\micron\ (J band), where future AO
systems should achieve good correction and which provides good angular
resolution. Observations at J~band are also interesting because of
their potential to descry the 1.27\,\micron\ O$_2$ band expected in
oxygen-rich planetary atmospheres
\citep[e.g.,][]{turnbull:2006,kawahara:2012}.  In terms of
astrophysical parameters, the baseline scenario assumes that
\verysmallp\ and \smallp\ planets occur with equal frequencies (as
described in Sec.~\ref{sec:pop}, this may be a conservative
underestimate), and that planet frequency is flat in $\ln P$ space for
$P>50$~d (the longest periods considered by \citealp{howard:2012}). 

\begin{figure}[tb!]
\centering
\includegraphics[width=\figwidth]{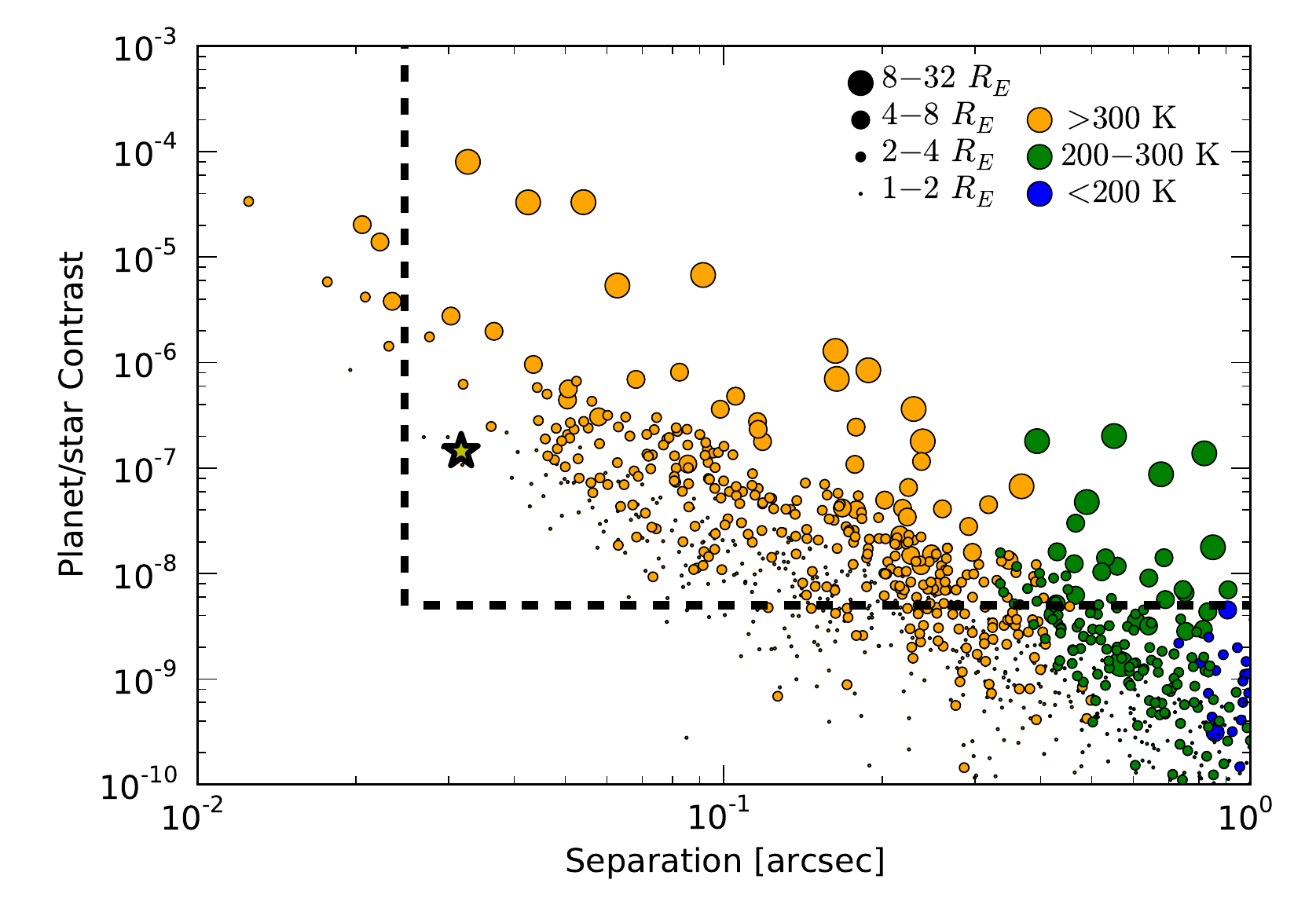}
\vspace{-0.3cm}
\caption{ \label{fig:obs1} Simulated planet populations for
  $\alpha$ Cen B in the baseline scenario outlined in
  Sec.~\ref{sec:baseline}, observing at 1.2\,\micron\ with an
  instrument similar to TMT/PFI or TMT/SEIT.  Planets above and to the
  right of the dashed line would be accessible to high-contrast
  spectroscopic characterization. The star symbol indicates
  $\alpha$~Cen~Bb (assuming a radius of 1.1\,\rearth). Each circle's
  size and color refers to a simulated planet's $R_P$ and $T_{\rm
    eq}$, as indicated in the legend. Most of the simulated planets
  shown are detected in scattered starlight, though thermal emission
  contributes to the shortest-period planets (including
  $\alpha$~Cen~Bb).}
\end{figure}

\begin{figure}[tb!]
\centering
\includegraphics[width=\figwidth]{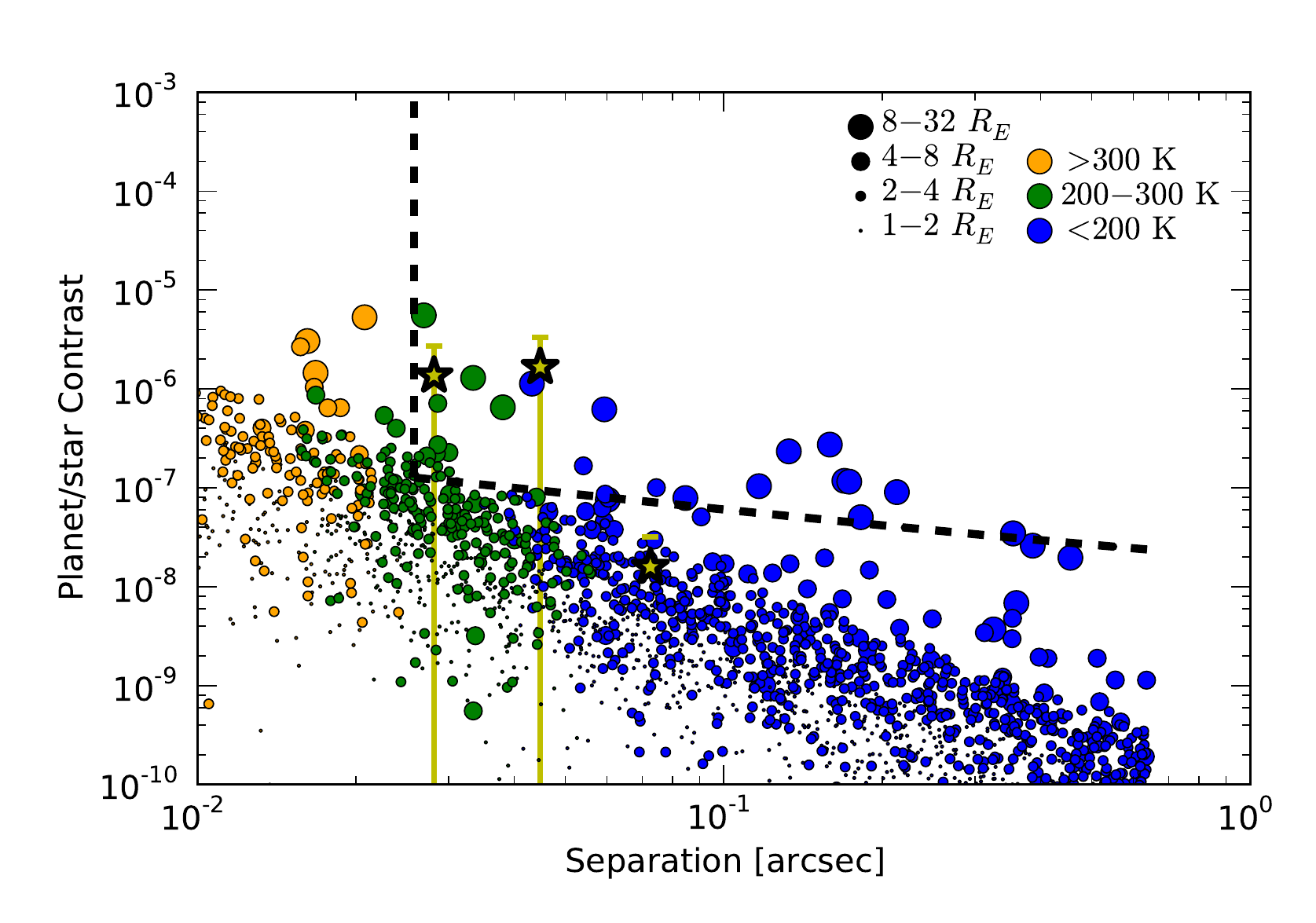}
\vspace{-0.3cm}
\caption{ \label{fig:gj876} Same as Fig.~\ref{fig:obs1}, but for the M
  dwarf GJ~876.  From left to right, the star symbols and error bars
  show the contrast expected for planets c, b, and~e over the
  range of albedos and recirculation parameters described in
  Sec.~\ref{sec:accessibility}. Planets~b and~c will be accessible to
  characterization in scattered light if $A_g (R_P/\rearth)^2\gtrsim$
  19~and 6, respectively.}
\end{figure}

Fig.~\ref{fig:obs1} shows a representative planet sample and the
computed sensitivity limits as applied to $\alpha$~Cen~B.  It is
notable that the recently discovered low-mass planet candidate
$\alpha$~Cen~Bb \citep{dumusque:2012} presents a contrast of $10^{-7}$
and is eminently observable in this baseline scenario \citep[assuming
a radius of $1.1\rearth$, obtained by converting $m\sin i$ to $R_P$
via Eq.~1 of ][]{lissauer:2011a}.  A simulation of the GJ~876 system
is shown in Fig.~\ref{fig:gj876}; despite the worse contrast achieved
on this fainter star, the massive planets GJ~876b and~c
\citep{delfosse:1998,marcy:1998,marcy:2001,baluev:2011} can be
directly characterized if $A_g (R_P/\rearth)^2\gtrsim$ 19~and 6,
respectively ($A_g\gtrsim0.02$, with radii estimated as described
previously).  Though not shown, the currently known planets GJ~139c
and~d \citep{pepe:2011} can also be characterized if $A_g
(R_P/\rearth)^2\gtrsim$ 0.9 and 1.8, respectively
(i.e. $A_g\gtrsim0.4$), and {the planet candidates $\tau$~Cet~b,
  c, and~d could be characterized for $A_g (R_P/\rearth)^2\gtrsim$
  0.08, 0.4, and 1.2, respectively ($A_g>$~0.04, 0.14, and
  0.35).} Theoretical efforts to model all these planets' NIR
reflectance are clearly warranted.

\begin{figure}[tb!]
\centering
\includegraphics[width=\figwidth]{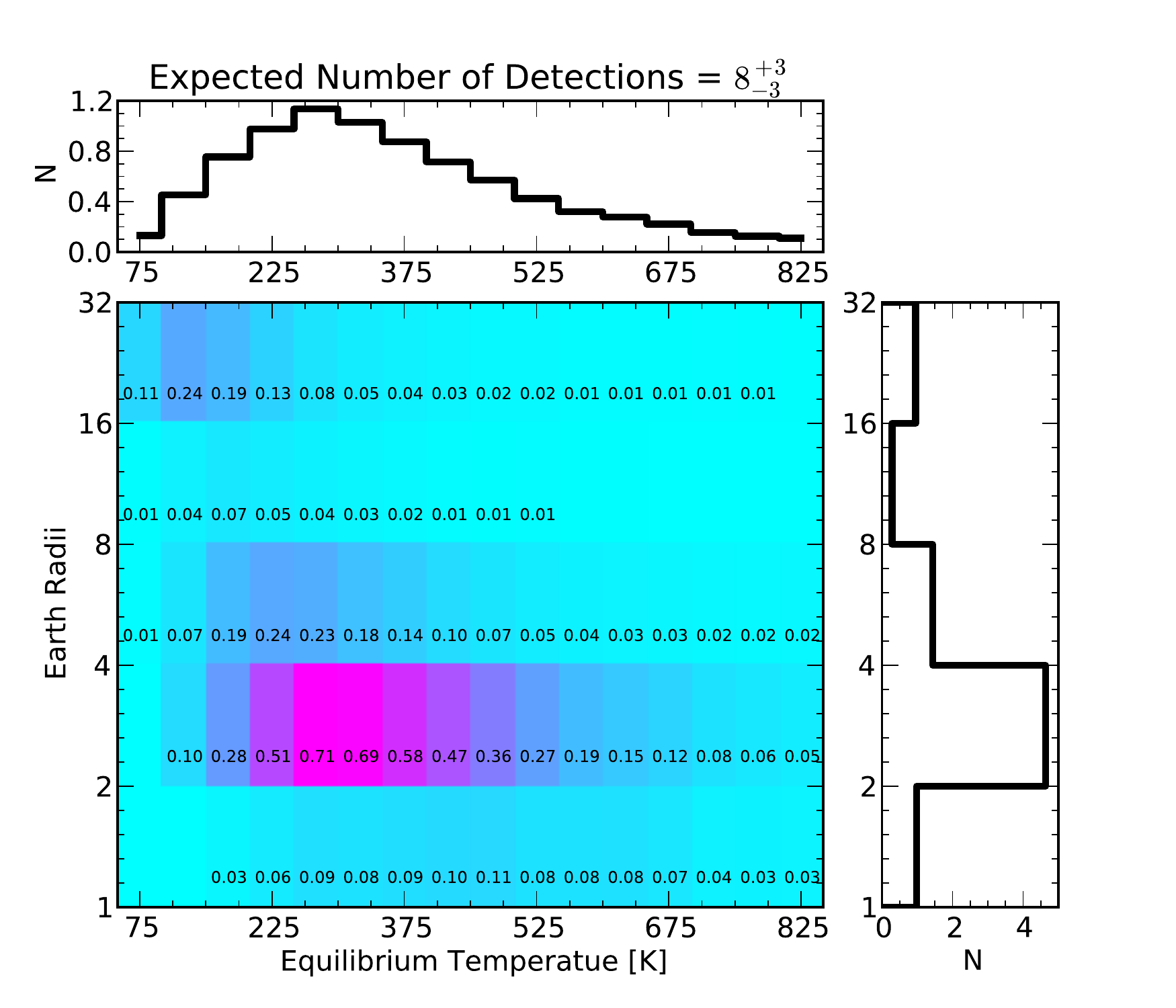}
\vspace{-0.3cm}
\caption{ \label{fig:hist1} Predicted number of planets (as a
  function of $R_P$ and $T_{\rm eq}$) expected to be characterized in
  the baseline high-contrast GSMT observing scenario outlined in
  Sec.~\ref{sec:baseline}.  The colors and the small numbers in each
  cell both indicate the expectation value for that combination of
  $R_P$ and $T_{\rm eq}$, and the histograms show the marginalized
  distributions. Note that the baseline scenario simulated here
  conservatively assumes that planets in the two smallest $R_P$ ranges
  occur with equal frequency (see Secs.~\ref{sec:pop}
  and~\ref{sec:astroparam}).}
\end{figure}

The number of accessible planets in this baseline scenario are shown
in Fig.~\ref{fig:hist1} as a function of $T_{\rm eq}$ and $R_P$, and
these results are summarized in the first row of
Table~\ref{tab:results}.  In all results I sort the planets by size
(in the ranges described previously) and according to whether $T_{\rm
  eq}$ is below, within, or above the range 200--300~K (a range which
corresponds crudely to the $T_{\rm eq}$ of Mars and a low-albedo
Venus).  Fig.~\ref{fig:hist1} shows that the planets detected in the
baseline scenario typically have $T_{\rm eq}<400$~K and
$R_P<4\,\rearth$: these planets are visible mainly via scattered
starlight and not intrinsic thermal emission.  Although
\cite{howard:2012} show that planets with $R_P<4\rearth$ are most
frequently found around cooler stars, I find that few planets orbiting
M dwarfs are accessible at distances $\gtrsim$3.5~pc as a result of
these stars' intrinsic faintness, and the consequent degradation of
achievable contrast. Finally, I estimate a 6\% chance that in this
baseline scenario an Earth/Venus analogue (\verysmallp, $T_{\rm
  eq}=200-250$~K) can be observed.  This probability is likely an
underestimate for the reasons described in Sec.~\ref{sec:pop}; such a
planet would most likely be found around Proxima~Centauri or Barnard's
Star. The probability increases to $>50\%$ when considering planets
with $R_P=1-4\rearth$; these temperate planets also most likely to be
found around nearby M dwarfs.

\subsection{Ability to Quantify the Local Exoplanetary Demographics}
\label{sec:astroparam}
Several assumptions were made in Sec.~\ref{sec:pop} to extrapolate the
Kepler population statistics of \cite{howard:2012} to longer orbital
periods and cooler stars, and in this section I test the effects of
 these assumptions. The results of these tests are all listed
in the second section of Table~\ref{tab:results}.

\cite{howard:2012} find roughly equal occurrence frequencies for
\verysmallp\ and \smallp\ planets. The baseline scenario described in
Sec.~\ref{sec:baseline} conservatively assumes that \smallp\ planets
and \verysmallp\ planets occur with equal frequency, {an
  assumption also consistent with more recent results
  \citep{dong:2013,fressin:2013}. For larger radii, \cite{howard:2012}
  find a planet frequency $\propto\, R_P^{-2}$ (around sun-like
  stars).  It now appears that this relation does not apply to
  $R_P<2\,\rearth$; nonetheless, Table~\ref{tab:results} shows that
  approximately equal numbers of planets in the two smallest size
  ranges would be accessible to characterization if the trend held to
  smaller radii \citep[i.e., if Kepler's completeness is significantly
  overestimated for
  $R_P<2\,\rearth$;][]{howard:2012,dong:2013,fressin:2013}.} However,
because the smaller planets must be closer to their host stars to
present the same contrast ratio as the larger planets, the additional
\verysmallp\ planets have rather warmer temperatures ($T_{\rm eq} \sim
400$~K) than the baseline population.

My baseline scenario conservatively assumes that planet frequency
flattens out for $P>50$~d, {consistent with recent results using
  additional Kepler data \citep{dong:2013}. A recent analysis of the
  first six quarters of Kepler data extends planet frequency estimates
  \citep{dong:2013}.  That work indicates that the frequency of
  planets in $\ln P$ space for $P<250$~d appears flat for
  $R_P<4\rearth$, but continues to increase for larger planets; this
  is consistent with radial velocity surveys, which show that the
  frequency of massive planets continues to increase out to 2000~d
  \citep{cumming:2008}.} I investigate the impact more frequent planet
occurrence on longer periods by letting \cite{howard:2012}'s planet
frequency relations increase at longer periods (instead of arbitrarily
flattening the distributions).  Table~\ref{tab:results} shows that
compared to the baseline scenario, two additional planets (one each
with sizes \mediump\ and \largep) would be accessible to high-contrast
characterization in this scenario. Because these additional planets
occur on longer periods, they are rather cooler ($T_{\rm eq}<200$~K)
than the baseline population.

Perhaps the least conservative assumption in my baseline scenario is
the decision to assign planets to stars with $T_{\rm eff}<3600$~K,
below the range explored by \cite{howard:2012}.  Large-scale radial
velocity campaigns have been conducted for a few such stars and have
found no planets down to $\sim10\mearth$ in short periods
\citep[e.g. Proxima Centauri and Barnard's Star;
][]{endl:2008,zechmeister:2009,choi:2012}.  Table~\ref{tab:results}
shows that even if all cool stars have {\em zero} planets, roughly
five planets would still remain for future studies.

{The immediately preceding assumption is of course overly
  pessimistic, since planets with $m \sin i>10\,\mearth$ have been
  discovered around a number of M dwarfs \citep[e.g.,][]{butler:2004,
    bonfils:2005, bonfils:2007,bonfils:2012}.  Furthermore, a
  systematic radial velocity survey of a large number of cool ($T_{\rm
    eff}<3600$~K) has recently indicated a planet occurrence frequency
  of roughly 90\%\ for $m \sin i=1-100\,\mearth$, $P<100$~d
  \citep{bonfils:2013}. Using the planet population statistics
  reported in Table~11 of that survey (instead of extending the
  \citealp{howard:2012} frequencies) for stars with $T_{\rm
    eff}<3600$~K, I predict numbers and types of planets amenable to
  high-contrast characterization nearly identical to that predicted by
  the baseline scenario (see Table~\ref{tab:results}). In this
  simulation I assume log-uniform frequencies of 2.7\% and 3\% for
  planets with $10^2-10^3$ and $10-10^2\,\mearth$, respectively, while
  the least massive planets ($1-10\,\mearth$) have frequencies of 36\%
  for $P=1-10$~d and 52\% for longer periods.  I convert $m \sin i$ to
  $R_P$ as in Sec.~\ref{sec:baseline} \citep{lissauer:2011a}, and
  impose a maximum radius of 1.3\,$R_J$. The close agreement between
  this simulation and the baseline scenario of Sec.~\ref{sec:baseline}
  hints that the results of \cite{howard:2012} and \cite{bonfils:2013}
  are consistent for small planets orbiting low-mass stars under
  fairly simple assumptions.  A conclusive test of such a claim is,
  however, beyond the scope of this work. }

\subsection{Effects of Instrumental Capabilities}
\label{sec:obsparam}
In this section I examine how the accessible exoplanet population
depends on various observational parameters: telescope diameter $D$,
inner working angle (IWA), and observing wavelength $\lambda$. The
results of all these tests are listed in the bottom section of
Table~\ref{tab:results}.

\subsubsection{Angular Resolution: Diameter and Inner Working Angle}
\label{sec:angular}
As discussed in Sec.~\ref{sec:ao}, several GSMT designs have been
proposed, each with different diameters $D$.  The baseline instrument
used in this work has an IWA defined as $3\lambda/D$, so the angular
magnitude of this IWA decreases as $D$ increases.  Planets at the
smallest separations have the highest observable contrast (in both
scattered light and thermal emission), so naturally
Table~\ref{tab:results} shows that the number of accessible planets
increases (by nearly a factor of two) when $D$ increases from 30\,m
(TMT) to 39\,m (E-ELT).  The number decreases by a similar factor when
$D$ is reduced to 25\,m (GSMT).

Table~\ref{tab:results} also shows that at most a single short-period
planet is predicted to be accessible to high-contrast instruments on
current 8\,m telescopes, {\em if} instruments on these telescopes can
reach contrast levels of $5\times10^{-9}$. In this case, the work
presented here predicts a $>1\%$ chance of accessible planets for only
four stars: $\alpha$~Cen~AB (15--20\% each), Procyon A (5\%), and
Sirius A (3\%). These expectation values drop by only a factor of 2 if
such facilities can only achieve contrasts of $10^{-8}$. These star
systems should therefore be high-priority targets for deep
observations in imminent 8\,m high-contrast surveys.

Similarly, using a narrower or wider IWA than the baseline
scenario of Sec.~\ref{sec:baseline} respectively increases or
decreases the number of accessible planets, as expected (see
Table~\ref{tab:results}).  However, varying IWA has a weaker impact
than varying $D$, even when the angular scale of the IWA is held
constant. This result may at first seem surprising: 2$\lambda/D$ on a
30\,m telescope gives a smaller angular separation than does
3$\lambda/D$ on a 39\,m telescope, yet more planets are accessible for
characterization under the latter conditions than under the
former. The increased sensitivity at larger $D$ for constant angular
IWA results from the $D^{-1}$ dependence of PSF speckle amplitude on
phase and amplitude errors \citep{guyon:2005}.  Thus for two
high-contrast instruments with equal IWA (in angular units), the
instrument with the larger primary mirror can reach better contrast
levels.

\subsubsection{Observing Wavelength: Scattered Starlight and Thermal Emission}
\label{sec:metis}
The choice of observing wavelength $\lambda$ has a number of important
consequences: with $D$ it sets the angular scale of the IWA (typically
specified in $\lambda/D$), it determines the stellar flux available
for sensing and correction of phase and amplitude errors, and it
modifies the planet/star flux contrast (in this work, by determining
the planet's thermal blackbody emission). Naturally, observations at
multiple infrared wavelengths are also desirable to observe spectral
features of different molecular species and/or surface compositions
\citep[e.g.,][]{kuiper:1947a,hu:2012}.

Table~\ref{tab:results} shows that the number of planets accessible to
high-contrast characterization drops steadily as the observing
wavelength increases from 1.0\,\micron\ to 3.5\,\micron. The detected
planets are cool and seen in reflective light, so their planet/star
contrast remains approximately constant even as angular resolution and
the photon flux available for AO wavefront sensing and correction both
decrease. Nonetheless, Table~\ref{tab:results} suggests that several
planets should exist which are accessible to characterization across
from at least $1-2.2\,\micron$. The number of Earth/Venus analogues
expected to accessible to characterization is a maximum at
1.6\,\micron, a value which presumably represents a tradeoff between
angular resolution and the stellar flux available for wavefront
sensing and correction.

Intriguingly, thermal-infrared observations appear to be more
attractive than observations at intermediate wavelengths such as
3.5\,\micron. All three GSMT projects have proposed AO-assisted MIR
instruments
\citep{brandl:2010,tokunaga:2010,okamoto:2010,hinz:2012}. The longer
wavelengths at which these instruments would operate roughly
correspond to the Wien peak of cool planets' blackbody emission,
thereby providing a method to measure radiometric radii
\citep{morrison:1973}, the surface compositions of airless planets
\citep{hu:2012}, and the potential to detect exoplanetary ozone
\citep[e.g.,][ and references therein]{rugheimer:2012}.

The MIR instrument simulated here is {\em not} conservative in one
respect: as in this work's other simulations, I assume that the
instrument's wavefront sensor and science instrument operate at the
same wavelength. Current designs for such instruments typically rely
on facility AO systems, whose wavefront sensors operate at shorter
wavelengths.  Under such a strategy, chromatic effects may degrade the
achievable contrast to the point that short-period planets become
inaccessible \citep{guyon:2005}. Instrument designers should therefore
consider the feasibility of MIR wavefront sensing for these
instruments, either with a separate MIR wavefront sensor or (to
minimize non-common wavefront errors) via phase retrieval and
correction using the main science camera
\citep[e.g.,][]{malbet:1994,borde:2006}.

\begin{figure}[tb!]
\centering
\includegraphics[width=\figwidth]{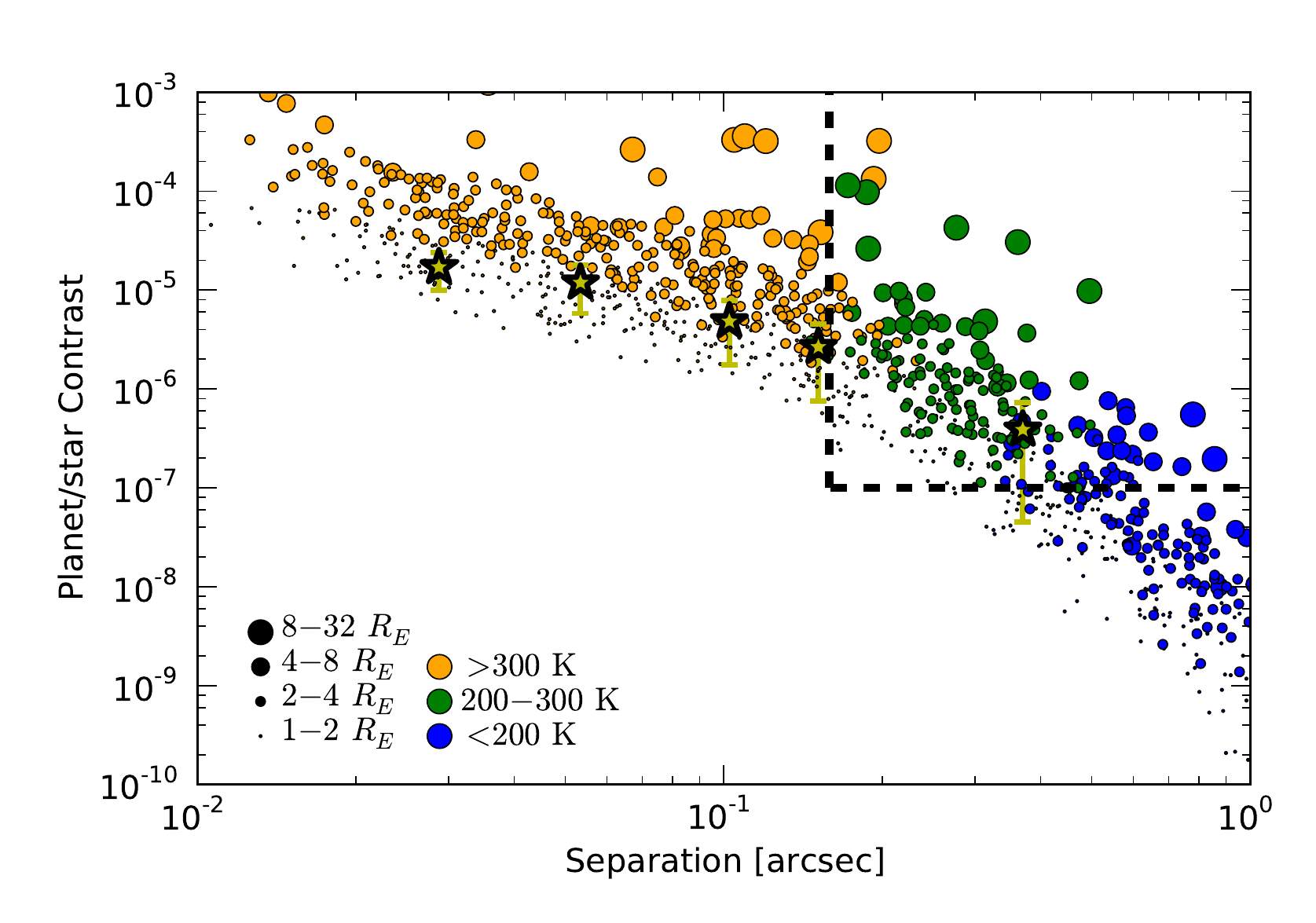}
\vspace{-0.3cm}
\caption{ \label{fig:obs2} Same as Fig.~\ref{fig:obs1} but for
  $\tau$~Cet in the thermal imaging scenario described in
  Sec.~\ref{sec:metis}, observing at 10\,\micron\ with an instrument
  similar to E-ELT/METIS.  The planets here are detected mainly via
  their thermal emission. From left to right, the star symbols and
  error bars show the contrast expected for planets candidates b,
  c, d, e, and~f \citep{tuomi:2013} over the range of albedos and
  recirculation parameters described in
  Sec.~\ref{sec:accessibility}. If confirmed, planet~f would be
  amenable to direct characterization of its thermal emission. }
\end{figure}

Fig.~\ref{fig:obs2} shows the planet population expected around
{$\tau$~Cet and the expected sensitivity limits for an E-ELT/METIS-like
instrument \citep[$\lambda=10\,\micron$, $D=39\,m$, and a conservative
minimum contrast of $10^{-7}$;][]{brandl:2012}.  Note that the planet
candidate $\tau$~Cet~f \citep{tuomi:2013} would be amenable to
high-contrast characterization at 10\,\micron, assuming a reasonable
albedo ($\lesssim 0.7$).}

\begin{figure}[tb!]
\centering
\includegraphics[width=\figwidth]{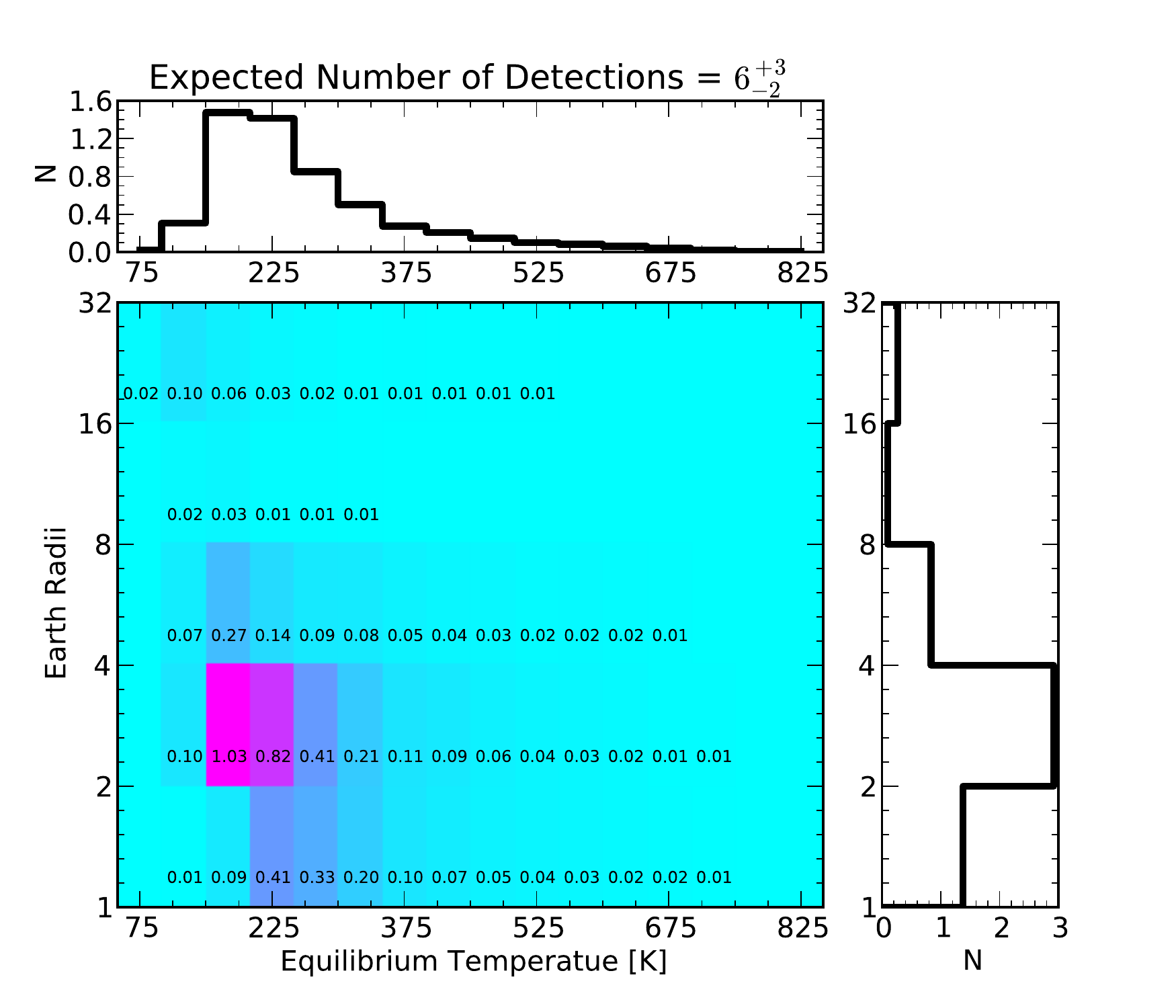}
\vspace{-0.3cm}
\caption{ \label{fig:hist2} Predicted number of planets (as a
  function of $R_P$ and $T_{\rm eq}$) expected to be characterized
  for the $10\,\micron$ imaging scenario described in Sec.~\ref{sec:metis}.
   The colors and the small numbers in each
  cell both indicate the expectation value for that combination of
  $R_P$ and $T_{\rm eq}$, and the histograms show the marginalized
  distributions. Compared to the planet population accessible in the baseline NIR
  case shown in Fig.~\ref{fig:hist1}, thermal infrared observations
  mainly detect thermal emission from cooler planets around warmer
  stars. Note that the  scenario simulated here conservatively
  assumes that planets in the two smallest $R_P$ ranges occur with
  equal frequency (see Secs.~\ref{sec:pop} and~\ref{sec:astroparam}).}
\end{figure}

A high-contrast MIR instrument has a much-reduced angular resolution
compared to NIR observations, but planetary thermal emission boosts
the planet/star contrast ratios significantly above the values seen at
shorter wavelengths; consequently, comparable numbers of planets are
expected to be accessible to NIR and MIR characterization.
Fig.~\ref{fig:hist2} shows the accessible local planet population as a
function of $R_P$ and $T_{\rm eq}$ for the thermal imaging simulation,
and the last row of Table~\ref{tab:results} summarizes these
results. 

High-contrast thermal-IR instruments thus offer the potential to
directly measure thermal emission from short-period planets down to
very small planetary radii. Thermal imaging shifts the distribution of
sampled planets to substantially cooler $T_{\rm eq}$ ($<200$~K) as
compared to the NIR baseline scenario of Sec.~\ref{sec:baseline}
(which is sensitive to rather hotter planets at smaller separations).
The number of \verysmallp\ planets accessible to thermal infrared
characterization is comparable to the number expected in the NIR
baseline scenario. However, even these conservative planetary
population statistics imply a 40\% chance of finding an Earth/Venus
analogue ($R_P=\verysmallp$, $T_{\rm eq}=200-250$~K); this probability
derives mainly from K-type stars with $T_{\rm eff}$ in the range
3900--5500~K, and so may be less sensitive to the assumptions made in
Sec.~\ref{sec:pop} for planet frequencies around cooler stars. As
discussed in Secs.~\ref{sec:pop} and~\ref{sec:astroparam}, if
\verysmallp\ planets are 4$\times$ more frequent than \smallp\ planets
then the expectation value increases to 1.6. Thus it may be considered
likely that such an object will be found; such an event would excite
considerable interest.

\section{Conclusions and Future Work}
\label{sec:conclusion}
These simulation of a mostly generic high-contrast instrument indicate
that, with reasonable extrapolations of planet frequency rates from
the Kepler mission and/or radial velocity surveys
\citep{howard:2012,bonfils:2013}, of order 10 short-period planets
will be accessible to characterization via high-contrast imaging and
spectroscopy from the next generation of ground-based giant
segmented-mirror telescopes (GSMTs). In Table~\ref{tab:results}, the
accessible planet population is summarized and shown to be insensitive
to moderate changes in the assumptions made about planet frequencies
(see Sec.~\ref{sec:astroparam}) and future instrumental capabilities
(see Sec.~\ref{sec:obsparam}).

Targets for future high-contrast characterization are already known
today. At wavelengths near 1.2\,\micron, the planets GJ~876b and~c
(see Fig.~\ref{fig:gj876}) and planet {candidates $\alpha$~Cen~Bb (see
Fig.~\ref{fig:obs1}) and $\tau$~Cet~b should be detectable to the NIR
baseline instrument considered in Sec.~\ref{sec:baseline}. GJ~139c
and~d and $\tau$~Cet~c and~d should also be detectable if their
geometric albedos are $>0.15-0.4$. Thus perhaps a half-dozen suitable
planets with a range of masses (and presumably radii) are already
known;} this number is large enough that one suspects the modest
predictions presented in this work may underestimate the true size of
the accessible exoplanet population. This could be the case if the
local exoplanet population differs substantially from that observed by
Kepler, or if the assumptions made in Sec.~\ref{sec:pop} are shown to
be invalid. Future high-precision radial velocity observations,
especially of M dwarfs, will enable further tests of these
predictions.

Thermal infrared observations will be desirable, because they offer
the possibility to directly characterize the thermal emission spectra
and energy budgets of these short-period planets, and thus (via
radiometric radii) break the radius-albedo degeneracy inherent in
reflected light observations. In particular, thermal infrared
observations significantly improve the chances of finding Earth or
Venus analogues (planets with $R_P=\verysmallp$ and $T_{\rm
  eq}=200-250$~K; see Table~\ref{tab:results}, and
cf. Figs.~\ref{fig:hist1} and~\ref{fig:hist2}). {The planet candidate
$\tau$~Cet~f (see Fig.~\ref{fig:obs2}) is the only currently known
planet which could be characterized by these mid-infrared
observations}, and radial velocity surveys seem likely to provide
additional targets in the coming decade.

The opportunity to observe and characterize the atmospheres of
relatively small ($R_P<4\,\rearth$) and cool ($T_{\rm eq}<300$~K)
planets is an exciting one, not least because such planets could be
similar to Earth and are particularly compelling objects in the study
of exoplanetary habitability.  A useful avenue for future study will
be to examine in more detail what discernible differences might be
used to distinguish Earth-like and Venus-like planets (which exhibit
similar infrared emission spectra) on the basis of mid-infrared
observations.  One initial avenue might be to extend planet
classification efforts using optical colors
\citep{crow:2011,hegde:2012} to longer wavelengths.

Future, more detailed simulations are required to estimate the signal
to noise with which the accessible planets discussed here can be
characterized. {\cite{kawahara:2012} have presented an excellent first
step toward this goal with their recent determination that GSMTs will
allow low-resolution spectroscopy of small planets at 1.27\,\micron\
in just a few hours. Further such work at both shorter and longer
wavelengths will help determine whether spectroscopy will be possible,
or whether characterization will be limited to broadband
photometry.} Such estimates would also permit predictions of the
precision with which high-contrast observations could measure
short-period planets' orbital properties (via astrometry), radiometric
radii (via MIR observations), and perhaps eventually measure rotation
periods \citep{ford:2001}, variations in climate \citep{cowan:2012b},
and perhaps exoplanetary satellites
\citep{robinson:2011,gomez:2012}. Future efforts would also be useful
to extend the work presented here to observations in polarized light
\citep{zugger:2010}. Finally, another interesting project would be an
assessment of the detectability of extrasolar comets
\citep{jura:2005}.

By the end of the current decade, JWST may obtain transmission spectra
of a few temperate, 1--3\,\rearth\ planets around M dwarfs
\citep{deming:2009,kaltenegger:2009}, but because transmission
spectroscopy probes the highest pressures such observations cannot
probe deeply into the atmosphere -- especially if opaque,
high-altitude hazes are as ubiquitous as they seem to be in the
atmospheres of hotter planets
\citep{bean:2011,berta:2012,pont:2012,benneke:2012}.  Nonetheless,
occultation measurements will be essential to inform models of
short-period planetary atmospheres and to place future high-contrast
GSMT observations in a broader context.

Future observational efforts are necessary to extend the planet
demographics measured by Kepler's initial planet candidate sample
\citep{howard:2012} to smaller planetary radii, cooler stars, and
(less importantly for this paper) longer orbital periods, {and indeed
such efforts are already being published
\citep{dong:2013,fressin:2013}.}  Focused high-precision radial
velocity campaigns of the type used to discover planet candidates
around $\alpha$~Cen~B and $\tau$~Cet \citep{dumusque:2012,tuomi:2013}
should be undertaken for the cool stars nearest the sun, both to
compare the frequency of planets in the solar neighborhood to that
observed by Kepler, and to discover additional planets for
high-contrast GSMT characterization.  High-precision spectra of all
stars within 8~pc of the Sun should also be obtained to characterize
their stellar properties, and thus better estimate the likelihood that
they host planets.

Planned GSMTs will have the potential to measure NIR albedos, MIR
emission, and consequently the molecular chemistry, energy budgets,
and radii, of significant numbers of small, temperate, and potentially
Earthlike planets around both M dwarfs and solar-type stars. Given
these positive prospects, continued efforts toward a more
comprehensive theoretical understanding of temperate exoplanet
atmospheres is certainly warranted.

\begin{acknowledgements}
  I thank B.~Macintosh and O.~Guyon for useful information, advice,
  and discussions while conducting this study, and the anonymous
  referee for useful comments. This research has made use of the
  Exoplanet Orbit Database at \url{http://www.exoplanets.org}, the
  Extrasolar Planet Encyclopedia Explorer at
  \url{http://www.exoplanet.eu}, and free and open-source software
  provided by the Python, SciPy, and Matplotlib communities.
\end{acknowledgements}


\clearpage

\begin{deluxetable}{l | c c c | c c c | c | c}
\tabletypesize{\small}
\tablecaption{Number of Short-Period Planets Accessible to High-Contrast Characterization \label{tab:results}}
\tablewidth{0in}
\tablehead{
\colhead{} & \multicolumn{3}{|c|}{Planet Size, $R_P/\rearth$} & \multicolumn{3}{|c|}{Equilibrium Temperature, $T_{\rm eq}$/K} & \colhead{} & \colhead{} \\
\colhead{Simulation Parameters} & \multicolumn{1}{|c}{$1-2$} & \colhead{$2-4$} & \colhead{$4-8$} & \multicolumn{1}{|c}{$<200$}& \colhead{$200-300$}& \colhead{$>300$} & \multicolumn{1}{|c}{$N_{\rm EVA}$ \tablenotemark{a}} & \colhead{Total\tablenotemark{b}} 
}
\startdata
Baseline Scenario (see Sec.~\ref{sec:baseline}):         & $1\pm 1$ & $5^{+2}_{-3}$ & $1^{+2}_{-1}$ & $2^{+2}_{-1}$ & $2^{+2}_{-1}$ & $4\pm 2$ & 0.06 & $8\pm 3$\\ 

\multicolumn{8}{l}{{\bf Effects of Astrophysical Parameters} (see Sec.~\ref{sec:astroparam}):}\\
Many Small Planets ($f_{\verysmallp} = 4 f_{\smallp}$) & $4\pm2$ & $5^{+2}_{-3}$ & $1^{+2}_{-1}$ & $3^{+1}_{-2}$ & $3^{+1}_{-2}$ & $6^{+2}_{-3}$ & 0.24 & $11^{+4}_{-3}$ \\ 
More Longer-Period Planets (for $P>50$\,d) & $1\pm 1$ & $5\pm 2$ & $2^{+2}_{-1}$ & $4\pm 2$ & $3^{+1}_{-2}$ & $4\pm 2$ & 0.06 & $10^{+4}_{-3}$  \\
No Planets Around Cool Stars ($T_{\rm eff}<3600$~K) &  $1\pm 1$ &  $3\pm 2$ &  $1\pm 1$ &  $0^{+1}_{-0}$ &  $1\pm 1$ &  $4^{+1}_{-2}$ &  zero & $5\pm 2$\\
RV Statistics for Cool Stars ($T_{\rm eff}<3600$~K) & $1^{+1}_{-1}$ & $4^{+1}_{-2}$ & $1\pm 1$ & $2^{+2}_{-1}$ & $2\pm 1$ & $4^{+1}_{-2}$ & 0.05 & $8\pm 3$ \\

\multicolumn{8}{l}{{\bf Effects of Instrumental Parameters} (see Sec.~\ref{sec:obsparam}):}\\
Larger Diameter (39\,m) &$2^{+2}_{-1}$ &  $9\pm 3$ &  $2^{+2}_{-1}$ &  $4\pm 2$ &  $4\pm 2$ &  $7^{+2}_{-3}$ &  0.14 & $15\pm 4$\\
Smaller Diameter (25\,m) & $0^{+1}_{-0}$ & $3^{+1}_{-2}$ & $1\pm 1$ & $1^{+2}_{-1}$ & $1\pm 1$ & $2^{+2}_{-1}$ & 0.04 & $5\pm 2$\\
Much Smaller Diameter (8\,m) &   0 & $0^{+1}_{-0}$ & 0 & 0 & 0 & $0^{+1}_{-0}$ & zero & $0^{+1}_{-0}$\\
Narrower IWA (2 $\lambda/D$)              & $1^{+2}_{-1}$ & $7^{+2}_{-3}$ & $2^{+1}_{-2}$ & $3^{+1}_{-2}$ & $3^{+1}_{-2}$ & $6^{+2}_{-3}$ & 0.07 & $11^{+4}_{-3}$ \\ 
Wider IWA (4 $\lambda/D$) & $1^{+0}_{-1}$ & $3^{+2}_{-1}$ & $1\pm 1$ & $2^{+2}_{-1}$ & $2^{+1}_{-2}$ & $3^{+1}_{-2}$ & 0.06 & $6^{+3}_{-2}$\\
Shorter Wavelength (1.0\,\micron)  & $1\pm 1$ & $5\pm 2$ & $1^{+2}_{-1}$ & $2^{+1}_{-1}$ & $2^{+2}_{-1}$ & $4^{+3}_{-2}$ & 0.05 & $9\pm 3$ \\ 
Longer Wavelength (1.6\,\micron)   & $1\pm 1$ & $4\pm 2$ & $1\pm 1$ & $2^{+2}_{-1}$ & $2^{+1}_{-2}$ & $3^{+1}_{-2}$ & 0.08 & $7\pm 3$ \\ 
Longer Wavelength (2.2\,\micron)   & $1^{+0}_{-1}$ & $2^{+2}_{-1}$ & $1\pm 1$ & $2^{+1}_{-2}$ & $1\pm 1$ & $2^{+1}_{-1}$ & 0.05 & $5\pm 2$ \\
Longer Wavelength (3.5\,\micron) & $1^{+0}_{-1}$ & $1^{+2}_{-1}$ & $1^{+0}_{-1}$ & $1\pm 1$ & $1^{+0}_{-1}$ & $2^{+2}_{-1}$ & zero & $3^{+2}_{-1}$\\
Thermal IR, Larger Diameter (10\,\micron, 39\,m, $10^{-7}$) & $1^{+2}_{-1}$ & $3\pm 2$ & $1\pm 1$ & $3\pm 2$ & $1\pm 1$ & $1\pm 1$ & 0.41 & $6^{+2}_{-3}$\\

\enddata
\tablenotetext{a}{Expected number of accessible Earth/Venus analogues
  ($R_P=1-2\rearth$, $T_{\rm eq}=200-250$~K), assuming planets with
  $R_P=\verysmallp$ and \smallp\ occur with the same frequency (see
  Secs.~\ref{sec:pop} and~\ref{sec:astroparam}).}
\tablenotetext{b}{Because of rounding, the sum of each row's values
  may not match the Total value listed.}
 \end{deluxetable}

\end{document}